\documentclass[twocolumn,prb,showpacs,amsmath,floatfix,unsortedaddress]{revtex4}
\usepackage{graphicx}

\begin{document}

\title{Pair distribution function and structure factor of spherical particles}

\author{Rafael C. Howell}
\email{rhowell@lanl.gov}
\affiliation{Materials Science and Technology Division, Los Alamos National Laboratory,
Los Alamos, New Mexico 87545, USA}

\author{Thomas Proffen}
\email{tproffen@lanl.gov}
\affiliation{Lujan Neutron Scattering Center, Los Alamos National Laboratory,
Los Alamos, New Mexico 87545, USA}

\author{Steven D. Conradson}
\email{conradson@lanl.gov}
\affiliation{Materials Science and Technology Division, Los Alamos National Laboratory,
Los Alamos, New Mexico 87545, USA}

\date{\today}

\begin{abstract}
The availability of neutron spallation-source instruments that provide total scattering powder
diffraction has led to an increased application of real-space structure analysis using the pair
distribution function.  Currently, the analytical treatment of finite size effects within pair
distribution refinement procedures is limited.  To that end,
an envelope function is derived which transforms the pair distribution function of an
infinite solid into that of a spherical particle with the same crystal structure.
Distributions of particle sizes are then considered, and the associated envelope function is used
to predict the particle size distribution of an experimental sample of gold nanoparticles from
its pair distribution function alone.  Finally, complementing the wealth of existing diffraction
analysis, the peak broadening
for the structure factor of spherical particles, expressed as a convolution derived from the
envelope functions, is calculated
exactly for all particle size distributions considered, and peak maxima, offsets, and asymmetries 
are discussed.

\centering LA-UR 05-8264
\end{abstract}


\maketitle

\section{Introduction}
There currently exists much scientific interest of the physical and chemical properties of
nanoparticles and nanodomains.  Generally investigations of particles with diameters of the order
of one micron are limited to powders, and from the diffraction data one performs Rietveld
refinement to extrapolate their structure and size, by using the Debye-Scherrer formula,
\cite{scherrer} extensions such as Debye function analysis,\cite{vogel} and small-angle
scattering.\cite{guinier}
Alternatively one derives the pair distribution function (PDF) from the diffraction
data,\cite{proffen} thus facilitating the study of size, correlated atomic motion,\cite{jeong}
short- to medium-range order,\cite{egami} and other
phenomena more apparent with a real-space treatment.  

A number of techniques have been developed to infer the size and structure of particles
from the PDF alone.  For example, the local structure has been determined by an interpretation of
the first peaks of the PDF,\cite{neder} and the size has been estimated from a Fourier transform of
the wide-angle Debye-Scherrer diffraction pattern.\cite{hall}   In this manuscript, a rigorous
approach to the determination of particle size is taken by rederiving the pair distribution
function of a single
spherical particle, expressed as an envelope function that multiplies the PDF of an
infinite crystal with the same crystal structure.\cite{germer}  A general class
of distributions of particle sizes is then considered, and an associated distributed envelope
function
is obtained.  Using experimental PDFs of both bulk gold and gold nanoparticles, calculated from
high-$Q$ neutron-powder-diffraction data, a distributed envelope function is used to
transform the PDF
of bulk gold to give a best fit replication of the PDF of gold nanoparticles.  Based on the
parameters of this envelope function,
the particle size distribution of the gold nanoparticles is predicted and compared to that
obtained experimentally.  Finally, a relationship
between the real-space envelope function and a $Q$-space
convolution function is established, the latter of which is to be applied to the structure factor
of an infinite crystal to obtain that 
of a distribution of spherical particle sizes.  The analytical form of the convolution function
allows for a quantitative analysis of Bragg peak maxima, widths, and asymmetries as a function
of peak position and particle size distribution.  A thorough account of the relationship between
the PDF and structure factor can be found in the literature.\cite{guinier, waser} 

The formalism used is identical to Peterson {\it et al.},\cite{peterson} with comparisons
to other definitions and nomenclature found in Keen.\cite{keen}
With little effort, the conclusions made here regarding the PDF and structure factor of spherical
particles can also be carried over to embedded spherical domains, with the stipulation that the
individual domains be uncorrelated with each other
(i.e., they have random orientations), and uncorrelated with the host matrix.  Indeed, diffraction
analysis using spherical geometries has already been succcessful in studies of water in mesopores
and micropores,\cite{steytler} as well as distributions of particle and void sizes in NMR
cryoporometry.\cite{webber, webber2}  The structure
factor analysis of spherical domains with well-defined atomic structures fits within the more
general context of disorder
within crystals, thus contributing to the analytical treatment of the associated diffuse
scattering.\cite{frey, ice, wagner}  It also provides a means of quantifying the diffraction limit
with respect to localized
lattice distortions, with spherical domains being a special case to be considered within the large
class of nanoscale heterogeneities already studied.\cite{garcia}

\section{The pair distribution function of a single spherical particle}

The microscopic pair density gives a distribution of atomic pair distances $r$ in a sample, 
weighted by the pair's scattering lengths
\begin{equation}
\rho(r)=\frac{1}{4\pi r^2 N}\sum_{i\neq j}\frac{b_ib_j}{\langle b\rangle^2}\delta(r-r_{ij}).
\label{rho}
\end{equation}
The following method of constructing $\rho(r)$ will be useful in determining its form with respect
to spherical particles.  For each $r$, define a spherical shell with this radius.  Let the center
of this
sphere coincide with the position of an atom $i$, and record as weighted $\delta$ functions every
atom $j$ that intersects the spherical shell.  Finally, divide the result by its surface area 
and the total number of atoms $N$.  The calculation of $\rho(r)$ for a spherical particle
limits the position
of the center of the sphere to the atoms within the particle itself, whereas the sphere's surface
can extend beyond.  Note that $\rho(r)$ for a single particle may only be a part of the total
microscopic pair density of a solid or solution.

The contribution to $\rho(r)$ of all atomic pairs $\{ij\}$ within a spherical particle is limited
to the
range $0<r_{ij}<2R$, where $R$ is its radius.  If the particle is in solution, or we consider an
ensemble of identical particles with random orientations embedded within a host lattice, then
$\rho(r)=\rho_0$ for $r>2R$, where $\rho_0$ is the constant atomic number density outside the
particle.  To simplify the following, let $\rho_0$ be equal to the number density of the particle
itself,
or take $\rho_0=0$ for empty space.  The essence of the problem addressed hereafter is to quantify
the relative
population of atomic pair distances between any two atoms within the particle and pair distances
where one atom resides outside of the particle.  That is, $\rho(r)$ will have $r$-dependent
contributions from both the microscopic pair density $\rho_c(r)$ of an infinite crystal
and the uncorrelated outside structure $\rho_0$,
\begin{equation}
\rho(r)=f_e(r,R)\rho_c(r)+\left[ 1-f_e(r,R)\right]\rho_0,
\label{rhos}
\end{equation}
where $0\leq f_e(r,R)\leq 1$.  When
$R\to\infty$, $f_e(r,R)=1$ for all $r$ and $\rho(r)$ is that of an infinite crystal.  When
$R\to 0$, $f_e(r,R)=0$ for all $r$ and $\rho(r)=\rho_0$.
$f_e(r,R)$ is the envelope function that we now derive when $R$ is between these limits.  

\begin{figure}[ht]
\centering
\includegraphics[width=3in]{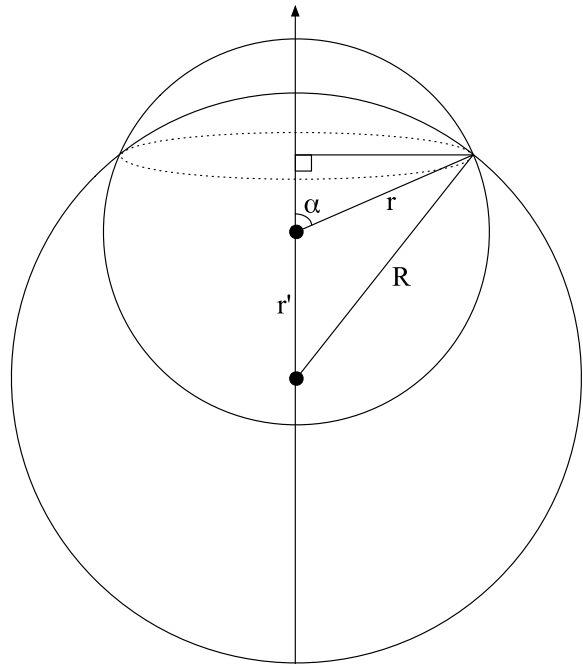}
\caption{A spherical particle with radius $R$.  An atom a distance $r'$ from the center of the
particle can have a shell of radius $r$ that is only partially embedded within the particle.}
\label{derivation}
\end{figure}

Consider a point within the particle whose position is given by the vector ${\bf r}'$, with the
center of the particle defining the origin.  Orient the point and the particle so that ${\bf r}'$
aligns with the $z$ axis, as shown in Fig. \ref{derivation}.  A spherical shell of radius $r$
around this point will either be enclosed by the particle if $r<R-r'$, intersect the surface of
the particle if $R-r'\leq r\leq R+r'$, or enclose the particle if $r>R+r'$.

When $R-r'\leq r\leq R+r'$, a line from any point on the circle of intersection and the position
$r'$ will meet the $z$ axis at an angle $\alpha$.  The fraction of the surface of radius $r$
around this point that is enclosed within the particle is
\begin{eqnarray}
f(r',r,R)&=&\frac{1}{4\pi r^2}\int_0^{2\pi}rd\phi\int_{\alpha}^{\pi}r\sin\theta\, d\theta \nonumber \\
 &=&\frac{1}{2}(1+\cos\alpha).
\label{surf}
\end{eqnarray}
Using the two right triangles from Fig. \ref{derivation}, the angle $\alpha$ can be expressed as
\begin{equation}
\cos\alpha = \frac{R^2-r'^2-r^2}{2r'r}.
\label{cos}
\end{equation}
The contribution of all such spheres enclosed within the particle is obtained by integrating
$f(r',r,R)$ over the remaining positions $r'$ in the region occupied by the particle,
taking care to consider when the shell of radius $r$ extends outside the region.  Using
Eq. \ref{cos}, for $r\leq 2R$,
\begin{eqnarray}
f(r,R)&=&\frac{4\pi}{3}(R-r)^3+4\pi\int_{R-r}^Rf(r',r,R)r'^2\, dr' \nonumber \\
&=&\frac{4\pi}{3}R^3\left[1-\frac{3}{4}\frac{r}{R}+\frac{1}{16}\left(\frac{r}{R}\right)^3\right].
\end{eqnarray}
Finally, dividing by the total particle volume gives the envelope function
\begin{equation}
f_e(r,d)=\left[1-\frac{3}{2}\frac{r}{d}
+\frac{1}{2}\left(\frac{r}{d}\right)^3\right]\Theta(d-r),
\label{envelope}
\end{equation}
where $d=2R$ is the particle diameter and $\Theta(x)= 0(1)$ for negative (positive) $x$ is the
Heaviside step function.  $f_e(r,d)$ and its derivative are continuous for all positive $r$.

The PDF of the particle is related to the microscopic pair density
\begin{eqnarray}
G(r,d)&=&4\pi r\left[\rho(r)-\rho_{\rm 0}\right] \nonumber \\
&=&f_e(r,d)G_c(r),
\label{pdf}
\end{eqnarray}
where $G_c(r)=4\pi r\left[\rho_c(r)-\rho_0\right]$ is the PDF of an
infinite crystal with the same crystal structure as the particle.

The fraction of atom pairs residing within a spherical region is obviously not a continuous
function of the region's size, considering the discrete nature of a crystal.  Therefore, one
expects the derivation above to become less accurate for smaller particle sizes.  To test this,
the exact calculation of $G(r)$ [using Eq. \ref{rho}] for spherical particles with an ideal fcc
structure with lattice constant $a$ was compared
to $G(r)$ obtained by applying Eq. \ref{pdf} to the PDF of an infinite ideal fcc structure.  A
particle was constructed with
$n$ shells of atoms around a central atom, giving it a diameter of $d=\sqrt{2n}\,a$ (some shells
contain no atoms).  While the difference in the nearest-neighbor peak heights between the two was
$7\%$ for $d=2a$, the difference between all peaks rapidly diminished to less than
$1\%$ for $d=4\sqrt{2}\,a$, and became negligible thereafter.

\section{Distributions of spherical particle sizes}

The envelope function $f_e(r,d)$ depends on a single-particle diameter $d$.
From a distribution $P(d')$ that defines an ensemble of particle diameters $d'$,
a distributed envelope function $f_{\rm DE}(r)$ can be constructed by weighting individual single
particle envelope functions with this distribution,
\begin{equation}
f_{\rm DE}(r)=\int_0^{\infty}f_e(r,d')P(d')\, dd'.
\label{prob_derive}
\end{equation}

Consider the following normalized distribution:
\begin{equation}
P(d',D,n)=\frac{1}{n!D}\left(\frac{d'}{D}\right)^n e^{-d'/D},
\label{prob}
\end{equation}
where $n$ is a positive integer and $D$ is a positive real number.
The average
particle diameter $d$ for this distribution is related to the parameters $n$ and $D$,
\begin{equation}
d=\langle d' \rangle=\int_0^{\infty}d'P(d')\, dd'=(n+1)D,
\label{d_ave}
\end{equation}
and the characteristic width of the distribution is 
\begin{equation}
\sigma=\sqrt{\langle d'^2 \rangle-d^2}=\frac{d}{\sqrt{n+1}}.
\end{equation}
Hereafter the average diameter $d$ will be used to identify the
associated distributed envelope functions and their behaviors, thus facilitating comparisons
between those obtained from various distributions and the single particle size envelope function.
The parameter $D$ is still used within function expressions, however,
to maintain their simplicity.  The two are always related by Eq. \ref{d_ave}.

When deriving a distributed envelope function from Eqs. \ref{prob_derive} and
\ref{prob} alone, one recognizes that a weighted sum of these distributions (each with unique
$n$ and $D$) yields a weighted sum of individual distributed envelope functions, so that
the results hereafter can be easily adapted to a variety of distributions.
For $n\geq 3$, a closed form expression of the resulting envelope function is given by
\begin{widetext}
\begin{equation}
f_{\rm DE}(r,d,n)=e^{-r/D}\sum_{k=0}^{n-2}\frac{1}{k!}\left(1-\frac{3}{2}\frac{k}{n}
+\frac{1}{2}\frac{k(k-1)(k-2)}{n(n-1)(n-2)}\right)\left(\frac{r}{D}\right)^k.
\label{dist_env}
\end{equation}
\end{widetext}
As with the single particle distribution $G(r,d,n)=f_{\rm DE}(r,d,n)G_c(r)$.

\begin{figure}[ht]
\centering
\includegraphics[width=3in]{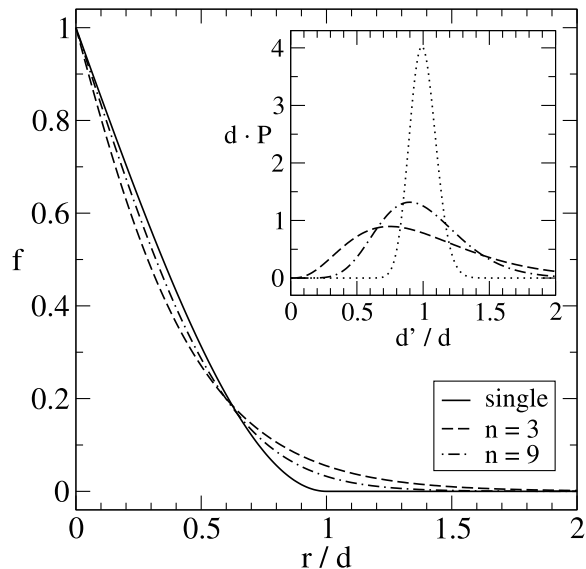}
\caption{The single particle size envelope function and two examples of
distributed envelope functions, with $n=3$ and $n=9$. The associated distributions $P(d'/d)$ are
shown in the inset along with $n=100$ (dotted).}
\label{env_dist}
\end{figure}

Figure \ref{env_dist} shows the single particle size envelope function $f_e(r,d)$
(Eq. \ref{envelope}) and the distributed envelope functions $f_{\rm DE}(r,d,n)$ for $n=3$ and
$n=9$ (expressed using the average diameters $d=4D$ and $d=10D$, respectively). The associated
distributions are shown in the inset, and the case of $n=100$ is included to
illustrate the trend of these distributions (the single particle size is a $\delta$ function
distribution, and is not shown). The width of the distributions clearly dictates the shape of
$f_{\rm DE}(r,d,n)$. A broader distribution gives the resulting envelope function a longer tail at
large $r$ at the expense of a steeper decline for small $r$.  Note that a
universal condition for {\it all} spherical particle size distributions with average
diameter $d$ is
\begin{equation}
f(r=0,d)=1\; \text{ and}\; \int_0^{\infty}f(r,d)\, dr=\frac{3}{8}d.
\label{identity}
\end{equation}

\section{Comparison with experiment}

By comparing an experimentally obtained PDF of a collection of spherical-like particles with the
theoretical expressions
just derived, it should be possible to predict the particle size distribution used
in the experiment.
Neutron data of a 2 g batch of capped gold nanoparticles
and a bulk gold fcc powder reference were collected at $T=15$ K on the
neutron powder diffractometer (NPDF) at the Lujan Center at Los Alamos National Laboratory.
\cite{page}  The PDF of both were obtained, and the bulk gold underwent full profile structural
refinements,
using PDFFIT,\cite{proffen2} to account for a host of influences such as correlated and
uncorrelated atomic motions and instrument resolution.  Multiplying the refined PDF of the bulk
data $G_{\rm b}(r)$ with an envelope function
derived from either a single particle size $f_e(r,d)$ or a distribution of particle sizes
$f_{\rm DE}(r,d,n)$ should give a good representation of the ideal PDF for spherical
nanoparticles of fcc gold.
A simplification is made by assuming that the crystal structure within the bulk gold and
gold nanoparticles is identical.  This simplification will be addressed when the results
of the PDF comparisons are discussed.

Using the PDF data of gold nanoparticles $G_{\rm np}(r)$ the best
fit diameter (and exponent for the distribution) is obtained by minimizing the root mean
square deviation $\delta_G$ between this PDF and the transformed bulk PDF.  The deviation is
given by
\begin{equation}
\delta_G^2= L^{-1}\int_{2.5}^{100.0}\left[ G_{\rm np}(r)-f(r,d)G_b(r)\right]^2\,dr,
\label{dev}
\end{equation}
where the lower bound is chosen to ignore spurious oscillations in the experimental PDF, the
upper bound is determined from the scattering resolution, and $L$ is the difference between these
two.  The integrations are performed numerically with a lattice spacing of $0.01$ \AA.

\begin{table}[ht]
\centering
\begin{ruledtabular}
\begin{tabular}{ l r r r r }
Distribution  & n  & d (\AA) & $\sigma$ (\AA) & $\delta_G$ (\AA$^{-2}$) \\
\hline
Single        &    & 28.75   &                & 1.243 \\
$P(d',d,n)$   & 13 & 29.80   & 7.96           & 1.238 \\
Experiment    &    & 35.48   & 13.12          & \\ 
\end{tabular}
\end{ruledtabular}
\caption{A comparison of the parameters from the best fit particle size distributions and the
experimentally determined distribution.}
\label{compare_table}
\end{table}

Table \ref{compare_table} compares the parameters from the best fit particle size distribution
and the single particle distribution. The predicted distribution of particle sizes,
with $n=13$
and $d_n=29.80$ \AA, gives the smallest absolute deviation from the experimental PDF, with
$\delta_G=1.238$ \AA$^{-2}$.

\begin{figure}[ht]
\centering
\includegraphics[width=3in]{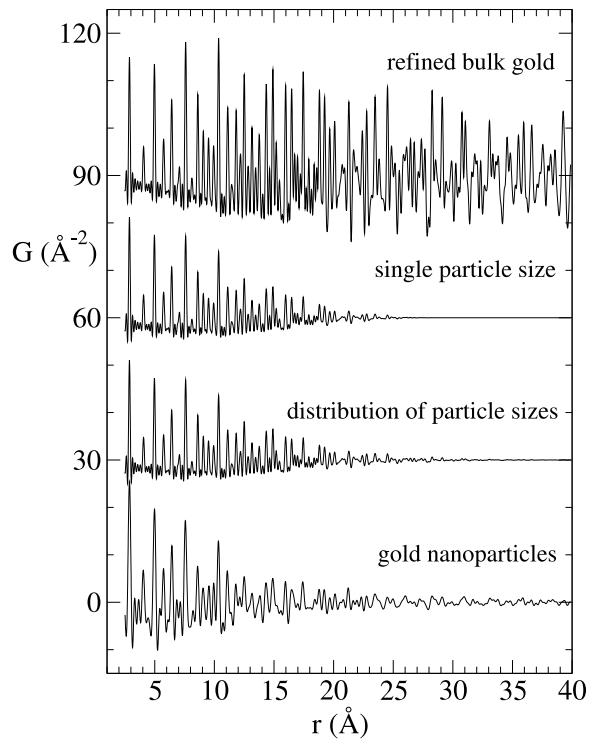}
\caption{An offset comparison of the refined experimental PDF from the bulk gold, the predicted
PDF from the single particle size and the distribution of particle sizes, and the experimental PDF
from the gold nanoparticles.}
\label{compare}
\end{figure}

Figure \ref{compare} shows an offset comparison between the experimental PDF of the gold
nanoparticles, the best fit PDFs from the single particle size and the distribution of particle
sizes, and the refined PDF of the bulk gold used to obtain the fits.  The envelope
function for the distribution of particle sizes allows for correlations over longer atomic
separations without changing the behavior of short-range correlations, as is evident in the figure,
and is thus a better fit to the experimental PDF than that of the single particle size distribution.
The PDF derived from a single envelope function always truncates to zero for $r\geq d$.

\begin{figure}[ht]
\centering
\includegraphics[width=3in]{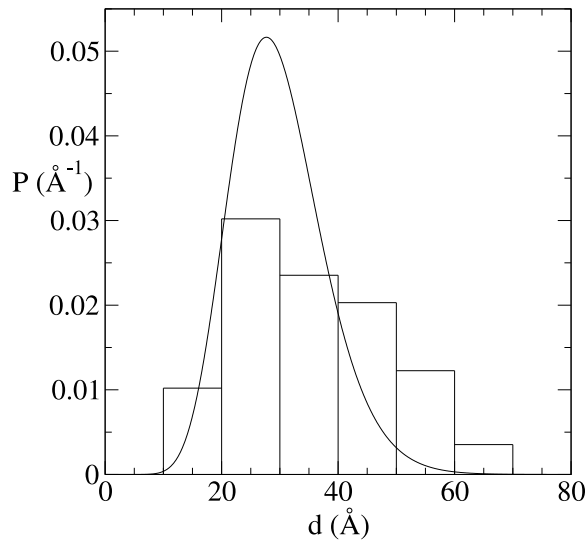}
\caption{A comparison of the experimentally determined particle size distribution
(histogram), and the predicted particle size distribution from the best fit PDF.}
\label{dist_fig}
\end{figure}

As a further test of the theory of spherical particles developed so
far, a comparison of the particle size distribution given by minimizing Eq. \ref{dev} with the
distribution obtained experimentally is now possible.  Using a JEOL 2010 TEM with point-to-point
resolution of $1.9$ \AA, the diameters of 148
gold nanoparticles within the experimental sample were measured, giving the distribution of
particle sizes shown in Fig. \ref{dist_fig}
(histogram).  The average particle diameter from this distribution is $d=35.48$ \AA, and the width
of the distribution is
$\sigma=13.12$ \AA. These values are given in Table \ref{compare_table}.  Also shown in the
figure is the
predicted theoretical particle size distribution
($d=29.80$ \AA, $\sigma=7.96$ \AA).  Obviously the best fit PDF predicts a distribution of
particle sizes
with a smaller average diameter and width. This can be attributed to the presence of pentagonal
twinning seen within the
gold nanoparticles (see Fig. 2 of Page {\it et al.}\cite{page}). The characteristic length scale
of the twinned regions defines the scale of
long-range order (an fcc structure in this case), which affects the resulting PDF.  Each object
might be better described as a collection of single domain crystallites instead of a spherical
particle with a  uniform structure.  Equation
\ref{dev} compares their PDF with the bulk gold PDF, in which twinning did not occur and the
structure was uniform, and thus
delivers a best fit envelope function that represents the size of the coherent fcc structure.  The
counting method used to construct the histogram in Fig. \ref{dist_fig}, however, considered only
total particle size, which contributes to the discrepancy between the two distributions.

\section{The structure factor of spherical particles}

The total-scattering structure factor of spherical particles $S(Q,d)$ can be obtained directly
from their PDF by a Sine transform
\begin{eqnarray}
Q\left[S(Q,d)-1\right]&=&\int_0^{\infty}G(r,d)\sin (Qr)\, dr \nonumber \\
&=&\int_0^{\infty}f(r,d)G_c(r)\sin (Qr)\, dr.
\label{sq}
\end{eqnarray}

Let $S_c(Q)$ be the structure factor of the associated infinite crystal. $f(r,d)$
is an envelope function from any distribution of particle sizes, and can be expressed as the
inverse cosine transform of a function ${\bar f}(Q,d)$,
\begin{equation}
f(r,d)=\frac{2}{\pi}\int_0^{\infty}{\bar f}(Q,d)\cos (Qr)\, dQ.
\end{equation}
Since $G_c(r)$ is the inverse Sine transform of $Q\left[S_c(Q)-1\right]$, Eq. \ref{sq}
can be written not only as a Sine transform of a product of two functions, but also as a
convolution of their respective transforms \cite{hildebrand}
\begin{eqnarray}
Q\left(S(Q,d)-1\right)&=&\frac{1}{\pi}\int_0^{\infty}\left[{\bar f}(|Q-Q'|,d)-
{\bar f}(Q+Q',d)\right] \nonumber \\
&&\times Q'\left(S_c(Q')-1\right)\, dQ' \nonumber
\end{eqnarray}
or
\begin{eqnarray}
S(Q,d)&=&\frac{1}{\pi Q}\int_0^{\infty}\left[{\bar f}(Q-Q',d)
-{\bar f}(Q+Q',d)\right] \nonumber \\
&&\times Q'S_c(Q')\, dQ'.
\label{conv}
\end{eqnarray}
The absolute value of the argument $Q-Q'$ can be disregarded if one recognizes the convolution
function ${\bar f}(Q,d)$ as an even function of $Q$.
 
For a single particle size distribution, the convolution function is
\begin{eqnarray}
{\bar f}_e(Q,d)&=&\int_0^{\infty}f_e(r,d)\cos (Qr)\, dr \nonumber \\
&=&\frac{3d}{(Qd)^2}\left(n_1(Qd)+\frac{1}{(Qd)^2}+\frac{1}{2}\right),
\end{eqnarray}
where $n_1(x)=-\cos(x)/x^2-\sin(x)/x$ is a spherical Bessel function of the second kind.
${\bar f}_e(Q,d)$ has a half-width at half-maximum of approximately $Q=3.48/d$.  This
convolution function is very similar to the expression used for the intensity of scattering from
randomly oriented identical spherical particles, and is often used in small-angle scattering
analysis,\cite{guinier} but differs from the sinc function derived from a Fresnel construction
of a spherical crystal that is often used to characterize powder diffraction lines.

The convolution function for the
distribution of particle sizes defined by Eq. \ref{prob} can be obtained by noting that each term
in Eq. \ref{dist_env} has a cosine transform proportional to
\begin{widetext}
\begin{equation}
\int_0^{\infty}\frac{1}{k!}\left(\frac{r}{D}\right)^ke^{-\frac{r}{D}}\cos(Qr)\, dr
=\frac{D}{\left[ 1+(QD)^2\right]^{k+1}}
\sum_{\substack{j=0\\ j \text{ even}}}^{k+1}(-1)^{j/2} \binom{k+1}{j} (QD)^j.
\end{equation}
\end{widetext}

\begin{figure}[ht]
\centering
\includegraphics[width=3in]{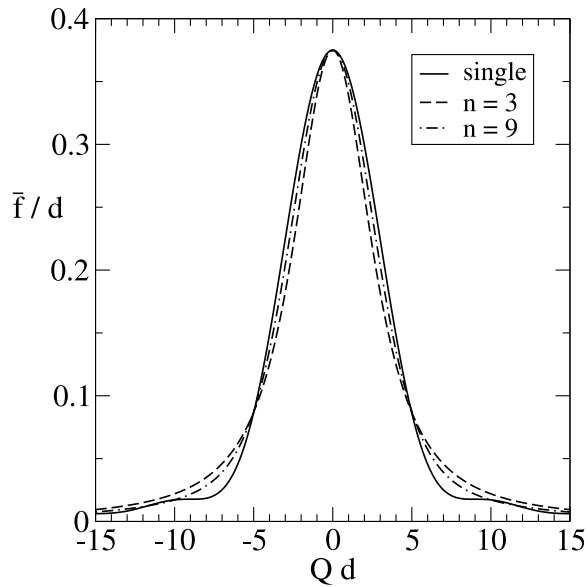}
\caption{The convolution functions to be applied to the infinite crystal structure factor
$QS_c(Q)$, for the single particle size and two distributions of sizes $n=3$ and $n=9$.}
\label{conv_pic}
\end{figure}

Figure \ref{conv_pic} shows the convolution functions for the single particle size and the two
distributions of sizes $n=3$ and $n=9$, considered before (expressed using the average diameters
$d=4D$ and $d=10D$, respectively).  For the single particle size,
shoulders appear after the primary peak, a feature attributed to the similarity between a periodic
envelope function and a triangular wave. When distributions of particle sizes are considered, the
shape is always Lorentzian-like, due to the exponential behavior of the associated distributed
envelope function.

A broader distribution gives the resulting convolution function a longer tail at large $Q$ at the
expense of a steeper decline for small $Q$.  Note that a corollary to Eq. \ref{identity} is
\begin{equation}
{\bar f}(Q=0,d)=\frac{3}{8}d\; \text{ and}\;\;
\frac{2}{\pi}\int_0^{\infty}{\bar f}(Q,d)\, dQ=1.
\end{equation}

To summarize, the structure factor of a distribution of particle sizes can obtained by convoluting
the structure factor of the associated infinite crystal with an envelope convolution function
${\bar f}(Q,d)$.  The result will be a broadening of the Bragg peaks of the infinite
crystal, but note that an asymmetry arises in this broadening, particularly for low $Q$, as
Eq. \ref{conv} is actually the difference of two convolutions, each weighted by $Q'$.

\begin{figure}[ht]
\centering
\includegraphics[width=3in]{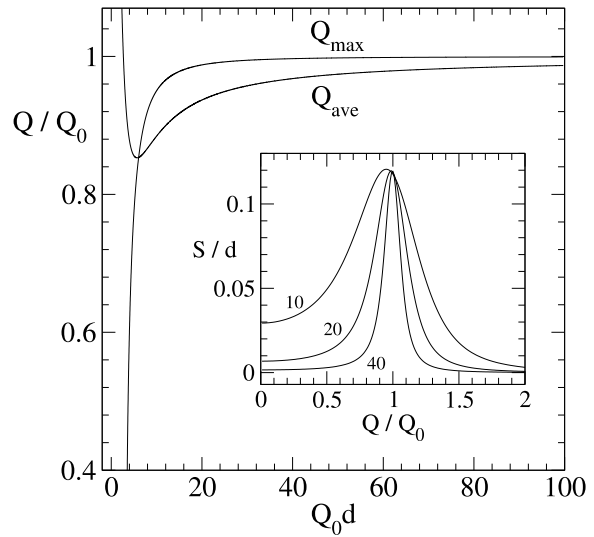}
\caption{A structure factor peak from spherical particles has a peak maximum and peak average
below the ideal peak position $Q_0$.  The asymmetries of the peak are illustrated in the inset for
$Q_0d=10$, $20$, and $40$.}
\label{conv_sq_pic}
\end{figure}

To illustrate this asymmetry, consider the effect that a convolution
resulting from finite particle sizes has on a single ideal Bragg peak from an
infinite crystal $S_c(Q)=\delta(Q-Q_0)$, with the effects on all other ideal Bragg peaks
being accounted for in a piecewise manner.  A
distribution of particle sizes with average
diameter $d$ and exponent $n=3$ gives a convolution function (shown in Fig. \ref{conv_pic})
\begin{equation}
{\bar f}_{\rm DE}(Q,d)=2d\frac{48+(Qd)^2}{\left[ 16+(Qd)^2\right]^2}.
\end{equation}
The convolution of the two [Eq. \ref{conv}] produces a broadened peak for $S(Q)$, shown in the
inset of Fig. \ref{conv_sq_pic} for $Q_0d=10$, $20$, and $40$.  A smaller average particle
diameter $d$
(or a lower value of the peak position $Q_0$) transforms the ideal peak of $S_c(Q)$ more
asymmetrically than a larger average particle diameter (or a higher peak position).  The peak
symmetry
is restored for very large values of $Q_0d$, as the subtracted term in Eq. \ref{conv} becomes
negligible.

Both the peak maximum $Q_{\rm max}$ and the average peak position (the normalized first moment)
$Q_{\rm ave}$ are also shown in the figure as a function of $Q_0d$.  For most values of $Q_0d$,
$Q_{\rm ave}<Q_{\rm max}<Q_0$, which suggests that not only does the peak position
shift to lower
values of $Q$, but it also diminishes slower to the left of the peak maximum than to the right.
For the extreme case when $Q_0d\lesssim 5.75$, $Q_{\rm max}$ quickly goes to zero, and as a
result $Q_{\rm ave}$
actually increases (it continues to broaden as $Q_0d$ decreases, but only for positive $Q$). This
occurs when the characteristic width of ${\bar f}_{\rm DE}(Q,d)$ is approximately $Q_0$.

\section{Conclusion}

The predictive power of having an analytical form of the PDF of spherical particles has been
clearly demonstrated.  In this case we were provided with structurally refined experimental PDF
data from a neutron source and a distribution of particle diameters observed from a TEM.  Only
the two together provided a means of testing the presented theory.
With care, one could arrive at the same predictions by deconvoluting the
experimental structure factor data of gold nanoparticles using Eq. \ref{conv}, but the
refinement analysis that afforded us a comparison between theory and experiment suggested
a real-space treatment.  Nevertheless, we believe that the convolution functions derived here
can complement the peak shape analysis already used in modern Rietveld packages,\cite{larson}
which often use Lorentzian and Gaussian functions alone to describe the peak broadening due to
particle size
effects.  Ideally, the analytical form of the peak shapes, as derived from the convolution
functions, should allow one to predict an
entire particle size distribution, and not just an average particle size, by considering together
the peak maxima, peak offsets, and peak asymmetries from an experimental structure factor.

It was mentioned earlier that the analysis of spherical particles should provide immediate insight
into the
effects of nanoscale domains embedded within a host lattice.  When two structures coexist as
uncorrelated domains, for example, a chemically disordered fcc structure and a second
chemically ordered structure with a small tetragonal distortion, the total PDF of the solid can
be taken as the sum of the individual PDFs, each calculated with a unique envelope function suitable
to their differing domain sizes.  The
broadening of the Bragg peaks resulting from their finite size, combined with the peak
splitting
of the second tetragonal phase, may dictate whether the chemically ordered domains are below
the diffraction limit, thus requiring a real-space probe such as XAFS to accurately measure their
local structure.
This analysis may be useful in explaining the presence of magnetism
in NiMn alloys when there is no signature of $L1_0$ ordering of the material apparent in the
diffraction data.\cite{espinosa}

Finally, the real-space treatment of PDF analysis, and the subsequent conclusions made about the
structure factor, is an encouraging approach to solving problems that are often only considered
in $Q$ space.  For example, the idea that a particle's surface might have a different structure
than its core (internal strain), can be realized in real space by considering both an envelope
function (for particle size), and a convolution function (for varying strain).  Can the
structure factor be derived from the two taken together, just as it was for the envelope function
alone, or does this problem require the $Q$-space treatments already considered?\cite{borbely}

\begin{acknowledgments}
This work was supported by the Heavy Element Chemistry Program, Chemical Sciences, Biosciences,
and Geosciences Division, Office of Basic Energy Sciences, and Defense Programs, NNSA, U.S.
Department of Energy under Contract No. W-7405.  It has also benefited from the use of NPDF at the
Lujan Center at Los Alamos Neutron Science
Center, funded by the DOE Office of Basic Energy Sciences, Los Alamos National Laboratory,
and the Department of Energy under Contract No. W-7405-ENG-36. The upgrade of NPDF has been funded
by NSF through Grant No. DMR 00-76488.
\end{acknowledgments}

\end{document}